\documentclass[nofootinbib,tightenlines,superscriptaddress,showpacs]{revtex4-2}
\usepackage{amssymb,amsmath,amsthm,graphicx,subfigure,float,cancel,braket,comment,tipa,bm,dsfont,graphicx,hyperref}
\usepackage{xcolor}
\usepackage{mathrsfs,enumitem,newtxtext} %newpxtext
\hypersetup{
	colorlinks,
	linkcolor={red!50!black},
	citecolor={blue!50!black},
	urlcolor={blue!50!black}
}
\usepackage{newtxmath}

\newcommand{\la}{\lambda}

\newcommand{\sig}{\sigma}
\newcommand{\bsig}{\bar{\sigma}}

\newcommand{\D}{\mathscr{D}}
\newcommand{\tf}{\frac{1}{2}}

\newcommand{\p}{\partial}
\newcommand{\al}{\alpha}
\newcommand{\be}{\beta}
\newcommand{\de}{\delta}
\newcommand{\ga}{\gamma}
\newcommand{\ep}{\epsilon}

\newcommand{\bet}{\bar{\eta}}

\newcommand{\bph}{\bar{\phi}}

\newcommand{\bze}{\bar{\zeta}}

\newcommand{\dal}{{\dot{\alpha}}}
\newcommand{\dbe}{{\dot{\beta}}}
\newcommand{\dga}{{\dot{\gamma}}}
\newcommand{\dde}{{\dot{\delta}}}

\begin{document}
\title{On FLRW reductions of N=1 supergravity}
\author{Nephtalí Eliceo Martínez Pérez}
\email{nephtali.martinezper@alumno.buap.mx}
\author{Cupatitzio Ramírez Romero}
\email{cramirez@fcfm.buap.mx}
\affiliation{Facultad de Ciencias Físico Matemáticas, Benemérita Universidad Autónoma de Puebla.}

\begin{abstract}
	An FLRW reduction of N=1 supergravity is troublesome since spatial isotropy prohibits a non-vanishing Rarita-Schwinger field. In view of this, we consider the quadratic form $\psi_m^{\ \al} \psi_{n\al}$ arising from a particular superspace generalization of the metric tensor. Imposing the FLRW form on this object reduces $\psi_m^{\ \al}$ to a single 2-component spinor $\phi^\al$, which becomes the superpartner of both the scale factor $a$ and lapse $N$, thus inducing the conformal time gauge, $N=a$. Time-reparametrization invariance is restored if we allow for deviations of FLRW form that are proportional to the gauge fields $N, \psi_0^{\ \al}$. We determine the allowed supergravity transformations that preserve the resulting FLRW supermultiplets.
\end{abstract}

\maketitle
\section{Introduction}\label{intro}
Friedmann-Lemaitre-Roberson-Walker or simply FLRW spacetimes are spatially homogeneous and isotropic, meaning they admit a six-dimensional group of isometries; three isotropies and three ``translations" acting on a special family of 3D spatial sections of constant time, which can be open, with zero or negative spatial curvature, or closed and positively curved \cite{ryanbook,inverno,ellis}. As a result of the symmetries, the dynamics of FLRW spacetimes resides in a single function, the \textit{scale factor} $a(t)$, accounting for the overall expansion of the spatial sections. A gauge field $N(t)$, called \textit{lapse}, also takes place in FLRW metric to make the time-reparametrization invariance manifest.

Dimensionally reduced FLRW, or Bianchi homogeneous models in general, are usually obtained by evaluating the 4D gravity action at the Bianchi metric and integrating over the spatial coordinates. The classical dynamics derived from the reduced actions is equivalent to that of the full theory, subject to the corresponding symmetries, whenever the Principle of Symmetric Criticality holds \cite{palais}. These 1D models are useful in quantum cosmology for they yield tractable versions of the quantum Hamiltonian constraint or Wheeler-DeWitt equation \cite{hartle,halliwell}.

The introduction of supergravity in quantum cosmology, called supersymmetric quantum cosmology, is relevant for several reasons \cite{deathbook,moniz}. In particular, the additional supersymmetric constraints of the corresponding canonical formulation provide a square root of the Hamiltonian constraint \cite{teitelboim,macias}. Upon quantization, this translates into more basic first-order equations that can change radically the space of physical states of the universe \cite{deathbook,moniz}.

Different strategies have been devised to construct supersymmetric cosmological FLRW models. On the one hand, the effective 1D FLRW action can be rendered supersymmetric by  introducing scalar Fermi degrees of freedom to complete  lapse and scale factor supermultiplets \cite{obregon1,obregon2,garcia,holten}. On the other hand, effective 1D actions with local supersymmetry have been obtained using an ansatz for the tetrad and Rarita-Schwinger fields that reduces their  number of independent components in a way consistent with local supersymmetry  \cite{death88,death92,asano,moniz,damour11,damour14}. In this work, we are interested in a supergravity generalization of FLRW spacetime by resorting to ideas of isometries and Killing vectors, just like it is done in Einstein's gravity and, if possible, make contact with some of the models described in the literature.

Killing vectors of spacetime have a natural generalization to Killing supervectors of curved superspace and its conformal generalization \cite{kuzenko,Kuzenko24}. (Conformal) Killing supervectors generate superspace transformations that, combined with certain Lorentz and Weyl transformations, determined by the Killing supervectors themselves, preserve the geometry of superspace, represented by the vielbein and spin connection or, equivalently, by the superspace covariant derivatives. Since supervectors parametrize both spacetime diffeomorphisms and gauged supersymmetry transformations, the concept of Killing vectors is naturally extended to Killing spinors.

The superspace approach constitutes a powerful method for constructing rigid supersymmetric field theories on (purely bosonic) curved backgrounds of arbitrary dimension, such as $AdS_4$, $\mathbb{R}\times S^3$, $S^4$, which admit a number of unbroken supersymmetries generated by Killing spinors \cite{Festuccia,Kuzenko13,Kuzenko25}. In these cases, the fermionic components of the super torsion and curvature vanish and, therefore, the Rarita-Schwinger field can be gauged away \cite{kuzenko2015}. 

FLRW backgrounds are not supersymmetric since the corresponding spinor Killing equations have no nontrivial solutions for generic $a(t)$. In this regard, investigations of gravitino phenomenology in the early universe are carried out in the context of field theory on curved spacetimes. The action of supergravity coupled to matter is linearized around a background comprising the FLRW metric, time-dependent scalar fields, and vanishing fermions. The mass term $m_{3/2}$ of the gravitino fluctuation is time-dependent because of the evolving background scalar fields. The effective dynamics is model dependent but some general features have been obtained, for a single chiral multiplet, the mass of transversal helicity-$\frac{3}{2}$ modes is essentially $m_{3/2}(t)$, whereas the effective mass of the longitudinal helicity-$\frac{1}{2}$ modes, arising through the super-Higgs mechanism, is more sensitive to the Hubble scale \cite{PhysRevD61,Kallosh2000,Giudice,Schenkel,Kolb}.

Now, in quantum cosmology the full gravitational and matter degrees of freedom are subject to quantization. In fact, only the homogeneous background fields are quantized in minisuperspace models \cite{hartle,Kiefer:2004xyv}. Supersymmetric cosmology requires a nontrivial FLRW background supermultiplet in order to realize a supersymmetric algebra on quantum cosmological states \cite{moniz,garcia}. For this purpose, the standard Killing equations prove inadequate as they yield a vanishing Rarita-Schwinger field, leaving no superpartner for the scale factor. Yet, as mentioned before, there is a handful of supersymmetric models, obtained by indirect means, whose bosonic sector corresponds to FLRW cosmology. 

In view of this, we consider a softened notion of isometries of N=1 supergravity in four dimensions, that is formulated in terms of the metric and the quadratic form $\psi_m^{\ \al} \psi_{n \al}$. The latter spinor bilinear emerges from a straightforward superspace generalization of the metric tensor. We use the superfield formulation of N=1 supergravity and the Wess-Zumino gauge. In this gauge, higher-$\theta$ components are used to carry the vielbein and spin connection to their simplest forms. This gauge still allows us to perform general coordinate, local supergravity and local Lorentz transformations. Sometimes this is also called Einstein's supergravity to distinguish it from the more general conformal supergravity \cite{kuzenko}.

We use two guiding principles for our FLRW reductions. First, the admission of continuous symmetries and their preservation under supergravity. This is established by the alternative Killing equations, complemented by a set of equations determining a subset of isometry-preserving supergravity transformations. Second, the emergence of a reduced supermultiplet with well-defined transformation rules. In general, symmetries eliminate a number of components of the metric and vector-spinor, but for this reduction to be useful in supersymmetric cosmology, the leftover components must constitute an autonomous  supermultiplet. This also requires that the eliminated components (expressed in a covariant way) do not reappear with a supergravity transformation. This last point is analogous to the spinor Killing equation, ensuring the preservation of $\psi_m=0$, with the difference that we deal with a non-vanishing Rarita-Schwinger field.

The paper is organized as follows. In Section 2, after a brief account of the superspace formulation of N=1 supergravity and the ordinary Killing equations in component form, we introduce quadratic Killing equations. In Section \ref{sec:2}, the k=0 and k=1 FLRW examples are worked out in detail. The content of this core section involves the following steps:
\begin{enumerate}
	\item Enforce the ordinary FLRW Killing vectors as solutions of the quadratic Killing equations, $\delta_X g_{mn}=0$, $\delta_X \psi_m \psi_n=0$, to determine the form of the metric and vector-spinor. We describe a working solution that keeps only the spin-$\frac{1}{2}$ component, $\phi^\al$, of the Rarita-Schwinger field, but removes the spin-$\frac{3}{2}$ component, $W_{\be \al \dga}$. This is consistent with the fact that the FLRW metric has no spin-2 component. Also, this contrasts with the free gravitino field on Minkowski spacetime, whose spin-$\frac{1}{2}$ component is constrained to vanish \cite{PhysRev.60.61,freedman}. 
	
	\item Find the subset of supergravity transformations that preserve the quadratic Killing equations by requiring $\delta_X (\delta_\zeta g_{mn})=0$, $\delta_X (\delta_\zeta \psi_m \psi_n)=0$. In this way, we find constraints that the isometry-preserving transformation parameters must satisfy. We consider in detail the cases with vanishing and positive spatial curvature. The latter example provides a generalization of the supersymmetric background $\mathbb{R} \times S^3$ \cite{Festuccia} to time-dependent radius.
	
	\item Isolate the FLRW supergravity multiplet and define its transformation rules. For a successful reduction to an FLRW supermultiplet, we impose by hand that $\delta_\zeta W_{\be \al \dga}=0$, which yields a final constraint on the transformation parameters. The FLRW reduction obtained in this way induces the conformal time gauge $N=a$, for which we call it conformal FLRW supergravity. 
	
	In a sense, steps 2 and 3 amount to find a sort of Killing spinors such that $\delta_\zeta W_{\be \al \dga}=0$, but that may well transform the spin-$\frac{1}{2}$ component of the Rarita-Schwinger field. 
\end{enumerate}
In Section \ref{sec:3}, we restrict the quadratic isometry equations to their spatial projection, which allows us to leave the gauge fields $N$ and $\psi_0$ arbitrary. This leads to a slight variation of FLRW supergravity of D'Eath and Hughes \cite{death88,death92}, hopefully in a more concise way, with the difference that we keep dotted/undotted fermions related by complex conjugation. Finally, Section \ref{sec:4} is dedicated to conclusions and outlook. Notation  and some handy formulas are collected in Appendix \ref{appendix1}. Appendix \ref{sphere} offers more details of the example of Section \ref{flat}.

\section{Supergravity and superspace}\label{sec:1}
For later reference, we recall here some basic ideas of the superspace formulation of N=1 supergravity in the Wess-Zumino gauge \cite{wessbagger,kuzenko}. Superspace is an extension of the spacetime manifold by anticommuting dimensions; a generic system of local superspace coordinates is  $z^M=(x^m,\theta^\mu,\bar{\theta}_{\dot{\mu}})$, where $x^m$ are ordinary spacetime coordinates and $\theta^\mu$, $\bar{\theta}^{\dot{\mu}}=(\theta^\mu)^*$ are anticommuting Grassmann coordinates with 2-component spinor indices. 

The geometry of superspace is provided by the vielbein and connection 1-forms, $\bm E^A=dz^M E_M^{\ A}(z)$ and $\bm{\phi}_B^{\ A}=dz^M \phi_{MB}^{\ \ \ \ A}(z)$ respectively. The internal super index $A=(a,\al,\dal)$ contains 4-vector and Weyl (dotted/undotted) Lorentz indices.

Under infinitesimal coordinate and local Lorentz transformations parametrized by $X^M(z)$ and $L_B^{\ A}(z)$, respectively, the vielbein transforms as follows \cite{wessbagger},
\begin{subequations}\label{basic}
	\begin{align}
		\delta_X E_M^{\ A}&=-X^L \partial_L E_M^{\ A}-(\partial_M X^L) E_L^{\ A}, \label{lorentz2} \\
		\delta_L E_M^{\ A}&=E_M^{\ B} L_B^{\ A} \label{lorentz3}.
	\end{align}
\end{subequations}

The covariant derivative of a Lorentz vector is defined with the spin-connection
\begin{align}
	\D_M X^A=\p_M X^A+(-)^{\ mb} X^B \phi_{MB}^{\ \ \ \ A}.
\end{align}
A combination of (\ref{lorentz2}) and the following field-dependent Lorentz transformation 
\begin{align}\label{super}
	K_B^{\ A}(X)\equiv -X^M \phi_{MB}^{\ \ \ \ A}
\end{align}
yields the covariant expression
\begin{align} \label{supergauge}
	\hat{\delta}_X E_M^{\ A}\equiv (\delta_X+\delta_{K(X)}) E_M^{\ A}=-\D_M X^A-X^B T_{BM}^{\ \ \ A}.
\end{align}
where the torsion components, $T_{BM}^{\ \ \ A}(z)$, are defined in (\ref{torsionr}).

The superfield components of the torsion and curvature are subject to the Bianchi identities \cite{wessbagger}. Still, they provide more field components than necessary for a consistent supergravity theory. Thus, some of them are eliminated by imposing superspace covariant constraints such as (\ref{const}) for the torsion components. 

Another way of eliminating superfluous components is by gauge fixing. Higher $\theta$-components of the transformation parameters $X^A(z)$ and $L_B^{\ A}(z)$ can be used to gauge away some components of the vielbein and connection \cite{ramirez}. In the Wess-Zumino (W-Z) gauge the lowest components of the vielbein and its inverse are given by
\begin{align}\label{wzgauge}
	E_M^{\ A}|_{\underline{\theta}=0} &=\begin{bmatrix}
		e_m^{\ a}  & \frac{1}{2} \psi_m^{\ \alpha} \\
		0 & \delta_\mu^{\ \alpha} 
	\end{bmatrix}, &&
	E_A^{\ M}|_{\underline{\theta}=0} =\begin{bmatrix}
		e_a^{\ m} & -\frac{1}{2} \psi_a^{\ \mu}  \\
		0 & \delta_\alpha^{\ \mu} \\
	\end{bmatrix},
\end{align}
with $\psi_a^{\ \mu}\equiv e_a^{\ n} \psi_n^{\ \alpha} \delta_\alpha^{\ \mu}$. 

Further, in the W-Z gauge, the Lie-algebra valued spin-connection is taken to the form
\begin{align}\label{wzcon}
	\big[\phi_{mB}^{\ \ \ A}, \phi_{\mu B}^{\ \ \ A}, \phi_{\dot{\mu} B}^{\ \ \ A}\big]_{\underline{\theta}=0}=\big[\phi_{mB}^{\ \ \ A}, 0 ,0\big]
\end{align}
The 4-vector $\phi_{mb}^{\ \ \ a}$ and spinor $\phi_{m\be}^{\ \ \ \al}$ representations are related by (\ref{gens}).

As in ordinary gravity, constraints (\ref{const}) allow us to solve for the connection in terms of the tetrad and vector-spinor. We use the following decomposition
\begin{align}
	\phi_{nml}\equiv e_m^{\ \ b} e_l^{\ a} \phi_{nba}=\omega_{nml}+\kappa_{nml},
\end{align}
where the bosonic connection with vanishing torsion $\omega_{nml}$, and the fermionic contribution $\kappa_{nml}$, called contorsion, are given in (\ref{spinconnection}). 
Thus, covariant derivatives take the form
\begin{subequations}\label{covz}
	\begin{align}
		\D_m X^a&=(\p_m X^a+X^c \omega_{mc}^{\ \ \ a})+X^c \kappa_{mc}^{\ \ \ a}\equiv D_m X^a+X^c \kappa_{mc}^{\ \ \ a}, \\
		\D_m \chi^\al&=(\p_m \chi^\al+\chi^\ga \omega_{m\ga}^{\ \ \ \al})+\chi^\ga \kappa_{m\ga}^{\ \ \ \al} \equiv D_m \chi^\al+\chi^\ga \kappa_{m\ga}^{\ \ \ \al}.
\end{align}
\end{subequations}

After fixing the W-Z gauge, we can still perform local Lorentz transformations arising from
\begin{align}\label{superlorentz}
L_B^{\ A}|_{\underline{\theta}=0}&\equiv \big(L_b^{\ a}, L_\be^{\ \al},L^\dbe_{\ \dal}\big),
\end{align}
plus spacetime diffeomorphisms and supergravity transformations arising from 
\begin{align}
	X^A|_{\underline{\theta}=0}& \equiv \big(X^a,\chi^\al,\bar{\chi}_\dal\big), \label{supervector}
\end{align}
(higher-$\theta$ components of $L_B^{\ A}$ and $X^A$ are determined from the preservation of the W-Z gauge \cite{wessbagger}).

Under a pure supergravity transformation arising from a supervector of the form $Z^A|=(0,\zeta^\alpha,\bar{\zeta}_{\dot{\alpha}})$, the tetrad and vector-spinor transform as  
\begin{subequations}\label{sugratrans}
	\begin{align}
		\delta_\zeta e_m^{\ a}&=i (\psi_m \sigma^a \bar{\zeta}-\zeta \sigma^a \bar{\psi}_m), \label{sugratetrad}\\
		\delta_\zeta \psi_m^{\ \alpha}&=-2 \mathscr{D}_m \zeta^\alpha+\frac{i}{3} (b_a \zeta \sig^a \bsig_m+3 b_m \zeta-M \bar{\zeta} \bsig_m)^\alpha \label{sugraspinor}
	\end{align}
\end{subequations}
where the real vector $b_m$ and the complex scalar $M$ are auxiliary fields.

\subsection{Isometries in gravity}
In the tetrad formulation of gravity, the metric is retrieved from the tetrad fields according to $g_{mn}=\eta_{ac} e_m^{\ a} e_n^{\ c}$, with the Minkowski metric $\eta_{ac}=\text{diag}(-1,1,1,1)$.

A spacetime vector $X^m(x)$ is a Killing vector if there is a local Lorentz generator $L_{ba}(x)$, depending both on $X^m$ and $e_m^{\ a}$, such that \cite{chinea,spinors} 
\begin{align}\label{isotetrad}
	\hat{\delta}_X e_m^{\ a}=-D_m X^a=e_m^{\ b} L_b^{\ a}.
\end{align}
(using the torsionless spin-connection (\ref{covz})). This condition ensures that the metric remains invariant under the one-parameter family of coordinate transformations generated by $X^m$, called isometries \cite{torres}. Indeed, contracting (\ref{isotetrad}) with $e_{na}$ and symmetrizing in $m, n$ yields
\begin{align}
\hat{\delta}_X g_{mn}&= -(D_m X^a) e_{na}-(D_n X^a) e_{ma} \equiv -D_m X_n-D_n X_m=0. \label{isometric}
\end{align}

Now, the infinitesimal transformation generated by another vector  $Y^m(x)$ is compatible with the isometries if
\begin{align}\label{preserve0}
	\hat{\delta}_X \hat{\delta}_Y g_{mn}&=\hat{\delta}_Y \hat{\delta}_X g_{mn}+\hat{\delta}_{[Y,X]} g_{mn}=\hat{\delta}_{[Y,X]} g_{mn}=0.
\end{align}
Therefore, $Y^m$ is either another Killing vector or an invariant vector field, $\delta_X Y=[Y,X]=0$. These restrictions on the allowed transformations follow from (\ref{preserve0}), which are complementary to the Killing equations (\ref{isotetrad}). In supergravity the parameters of gauged supersymmetric transformations will also be restricted to preserve the isometries.

\subsection{Isometries in supergravity}
A straightforward supergravity extension of the Killing equations follows from superspace generalization of (\ref{isotetrad}), 
\begin{align}\label{superiso}
	\hat{\delta}_X E_M^{\ A}=-\D_M X^A-X^B T_{BM}^{\ \ \ A}=E_M^{\ B} L_B^{\ A},
\end{align}
using definition (\ref{supergauge}). See \cite{kuzenko,Kuzenko24} for an exhaustive treatment of (conformal) Killing vectors of superspace.

Evaluating the lowest $M=m$ components of (\ref{superiso}) in the W-Z gauge (\ref{wzgauge}), we obtain,
\begin{subequations}\label{sugrakilling}
	\begin{align}
		\hat{\delta}_X e_m^{\ a}&=-\D_m X^a-X^B| T_{B m}^{\ \ \ a}|=e_m^{\ b} L_b^{\ a},  \label{tetradqf} \\
		\hat{\delta}_X \psi_m^{\ \al}&=-2 \D_m \chi^\al-2 X^B| T_{Bm}^{\ \ \ \al}|=\psi_m^{\ \beta} L_\beta^{\ \alpha}, \label{spinorvector}
	\end{align}
\end{subequations}
Unlike the non-supersymmetric case (\ref{isotetrad}), we now have some non-vanishing torsion components given in (\ref{torsion}).

Killing spinors are thus naturally incorporated into the theory: Replacing $X^A| \to Z^A|=(0, \zeta^\al, \bar{\zeta}_\dal)$, $\hat{\delta}_X$ becomes a supergravity transformation \cite{wessbagger}. For bosonic backgrounds, $\psi_m=0$, (\ref{tetradqf}) holds automatically, that is, $\delta_\zeta e_m^{\ a}=0$, (with $L_b^{\ a}=0$), whereas (\ref{spinorvector}) becomes the spinor Killing equation, $\delta_\zeta \psi_m^{\ \al}=0$.

In this work, we focus on ordinary coordinates transformations generated by a spacetime vector $X^m(x)$ or, embedding it into a superspace vector,
\begin{align}\label{ordinary}
	X^M|_{\underline{\theta}=0}=(X^m(x),0,0)
\end{align}
The corresponding Lorentz supervector is obtained by means of the vielbein, $X^A\equiv X^M E_M^{\ A}$. Using (\ref{wzgauge}) we obtain the field-dependent parameters
\begin{subequations}\label{diffeo}
\begin{align}
	X^a(x)&\equiv (X^M E_M^{\ a})|_{\underline{\theta}=0}=X^m(x) e_m^{\ a}(x), \\
	\chi^\al(x)&\equiv (X^M E_M^{\ \al})|_{\underline{\theta}=0}=\frac{1}{2} X^m(x) \psi_m^{\ \al}(x).
\end{align}
\end{subequations}

Evaluating (\ref{sugrakilling}) with the field-dependent parameters (\ref{diffeo}), using the torsion components (\ref{torsion}), and making the following redefinition $L_b^{\ a} \to L_b^{\ a}+X^n \phi_{n b}^{\ \ \ a}$,  we obtain the Killing equations in terms of the spacetime vector $X^m$
\begin{subequations}\label{killingordinary}
	\begin{align}
	\delta_X e_m^{\ a} &\equiv -X^l \p_l e_m^{\ a}-(\p_m X^l) e_l^{\ a}=e_m^{\ b} L_b^{\ a}, \\
	\delta_X  \psi_m^{\ \al} &\equiv -X^l \p_l \psi_m^{\ \al}-(\p_m X^l) \psi_l^{\ \al}=\psi_m^{\ \beta} L_\beta^{\ \alpha}. \label{spinvec}
	\end{align}
\end{subequations}

If the solution of (\ref{killingordinary}) has non-vanishing vector-spinor, the preservation of isometries under supergravity transformations, namely,
\begin{subequations}\label{pres}
\begin{align}
	\delta_X [\delta_\zeta e_m^{\ a}]&=(\delta_\zeta e_m^{\ b}) L_b^{\ a}, \label{prestetrad}\\
	\delta_X [\delta_\zeta \psi_m^{\ \al}]&=(\delta_\zeta \psi_m^{\ \beta}) L_\be^{\ \al}, \label{presvectorspin}
\end{align}
\end{subequations}
yields the following constraints $\delta_X \zeta\equiv -X^l \p_l \zeta=\delta_L \zeta$ and $\delta_X \phi_{mb}^{\ \ \ a}=\delta_L \phi_{mb}^{\ \ \ a}$. The latter holds automatically when the connection is expressed in terms of the vielbein components.

Let's consider the k=0 FLRW example. Spatial homogeneity yields $e_m^{\ a}=e_m^{\ a}(t)$ and $\psi_m^{\ \al}=\psi_m^{\ \al}(t)$, in which case $L_b^{\ a}=0$ (spatial dependence may enter through a local Lorentz transformation). Thus, allowed supergravity transformations depend only on time, $\zeta=\zeta(t)$. Going ahead with spatial isotropy, (\ref{spinvec}) yields $\psi_m=0$. Now, (\ref{prestetrad}) vanishes automatically, but (\ref{presvectorspin}) does not. Thus, we impose the Killing spinor equation $\delta_\zeta \psi_m^{\ \alpha}=0$. Omitting auxiliary fields, we obtain,
\begin{align}
	\mathscr{D}_0 \zeta^\alpha=\partial_0 \zeta^\alpha=0, && 
	\mathscr{D}_i \zeta^\alpha=\frac{\dot{a}}{a} (\zeta \sigma_i \bar{\sigma}^0)^\alpha=0,
\end{align}
which do not have nontrivial solutions for a generic scale factor.

The situation changes in the relaxed scenario considered below in that $\psi_m\ne 0$, and the Killing spinor equation is not invoked.

\subsection{Quadratic Killing equations}
Lorentz generators can be eliminated of the linear Killing equations by contracting (\ref{tetradqf}) and (\ref{spinorvector}) with $e_{na}$ and $\psi_{n\al}$, respectively, and symmetrizing in $m, n$,
\begin{subequations}\label{quadratic0}
\begin{align}
\hat{\delta}_X g_{mn}&=-\D_{(m} X^a e_{n)a}-X^B T_{B(m}^{\ \ \ a} e_{n)a}|=0, \label{killmet} \\
\hat{\delta}_X \psi_m \psi_n&=-\D_{(m} \chi^\al \psi_{n)\al}-X^B T_{B(m}^{\ \ \ \al} \psi_{n)\al}|=0. \label{quad}
\end{align}
\end{subequations}
where $\psi_m \psi_n\equiv \psi_m^{\ \al} \psi_{n \al}=\psi_n^{\ \al} \psi_{m \al}$ is symmetric due to the anticommuting nature of the spinor components.

Using higher $\theta$-components of $X^A$ \cite{wessbagger}, (\ref{killmet}) can be written as  $-e_{(n}^{\ a} e_{m)}^{\ b} \D_b X^a|=0$, which resembles (\ref{isometric}) (cf. \cite{kuzenko}).

It can be seen that equations (\ref{quadratic0}) establish the independent vanishing of the bosonic and fermionic parts of a single expression, namely, $\hat{\delta}_X (E_m^{\ A} E_{nA})|=0$, where 
\begin{align}\label{supermetric}
	(E_m^{\ A} E_{nA})|=e_m^{\ a} e_{na}+\frac{1}{4} (\psi_m^{\ \al} \psi_{n\al}+\bar{\psi}_{m\dal} \bar{\psi}_n^{\ \dal})
\end{align}
The real symmetric tensor sitting next to the metric defines a ten-parameter equivalence class of vector-spinors related by local Lorentz transformations, in the same way as the metric amounts to a ten-parameter equivalence class of tetrads.

Since equations (\ref{quadratic0}) are Lorentz invariant, we can replace the supergauge transformation $\hat{\delta}_X$ with the ordinary coordinate transformation, $\delta_X$, given by (minus) the Lie derivative \cite{torres},
\begin{subequations}\label{quadratic1}
	\begin{align}
		\delta_X\, g_{mn}&\equiv -\mathcal{L}_X\, g_{mn}=0, \label{iso}\\
		\delta_X \, \psi_m \psi_n&\equiv -\mathcal{L}_X\, \psi_m \psi_n=0. \label{quadratic}
	\end{align}
\end{subequations}
where $\psi_m \psi_n\equiv \ep_{\al \ga} \psi_m^{\ \al} \psi_n^{\ \ga}=\psi_n \psi_m$. We are thus imposing the vanishing Lie derivative of the complex $\psi_m \psi_n$, not only the real component suggested by (\ref{supermetric}).

Whereas (\ref{iso}) is equivalent to (\ref{tetradqf}), (\ref{quadratic}) is a little more general than (\ref{spinorvector}) since the former allows a non-vanishing homogeneous and isotropic solution as we shall show below.

In analogy with (\ref{preserve0}), the subset of supergravity transformations that preserve (\ref{quadratic1}) is found by requiring that
\begin{subequations}\label{superpres}
	\begin{align}
	\delta_X [\delta_\zeta\, g_{mn}]&=0, \label{preserve}\\
	\delta_X [\delta_\zeta\, \psi_m \psi_n]&=0, \label{preserve2}
	\end{align} 	
\end{subequations}
which will lead to constraints for the supergravity transformation parameters.

Equations (\ref{quadratic1}) and (\ref{superpres}) correspond to the first guiding principle mentioned above. The second principle will be addressed in Section \ref{supermiltiplet}, after solving the quadratic Killing equations.

\section{FLRW supergravity}\label{sec:2}
The Killing vectors of k=0 FLRW spacetime are, in Cartesian coordinates, $\bm T_i=\p_i$ and $\bm R_i=\ep_{ijk} x^j \p_k$, which generate spatial translations and rotations, respectively. Enforcing them on (\ref{quadratic}) leads to the following form
\begin{align}\label{gfrw}
	(\psi_m \psi_n)^{\text{FLRW}}&=\begin{bmatrix}
		(\psi_0 \psi_0) (t) & 0 \\
		0 & (\psi_1 \psi_1) (t) \delta_{ij}
	\end{bmatrix}.
\end{align}
The vanishing of the spatial vector $\psi_0 \psi_i$ relates the otherwise arbitrary gauge field $\psi_0$ to  the spatial components $\psi_i$. Assuming $\psi_0\ne 0$, the general solution of $\psi_0 \psi_i=0$ is $\psi_{i \al}=C_{i\al \ga} \psi_0^{\ \ga}$ for a traceless $(C_i)_{\al \ga}$. In a certain frame of reference, we may choose them proportional to the Pauli matrices, more precisely, we set
\begin{align}\label{ans}
	\psi_{k \al}=A_k (\sig_k \bsig^0 \psi_0)_\al, && \bar{\psi}_{k \dal}=A_k^* (\bar{\psi}_0 \bsig^0 \sig_k)_\dal 
\end{align}
with complex $A_k(t)$, for then $\psi_0 \psi_k=-\psi_0 \psi_0 A_k \delta_k^{\ 0}=0$ and $\psi_i \psi_k=A_i A_k g_{ik} \psi_0 \psi_0$.  Next, the equality $\psi_1 \psi_1=\psi_2 \psi_2=\psi_3 \psi_3$ leaves only one independent function, say $A_1(t)$.

Therefore, imposing spatial homogeneity and isotropy on $\psi_m \psi_n$ leaves us with the six real components contained in $\psi_0$ and the complex function $A_1(t)$, something like having one a half Weyl spinors. Now, we proceed to study the conservation of (\ref{quadratic1}) under supergravity transformations.

\subsection{Compatible supergravity transformations}
Instead of expanding (\ref{superpres}) in the general case, as in (\ref{preserve0}) (a task amenable to the superfield formulation), we compute directly the left-hand sides of (\ref{superpres}) using the transformation rules (\ref{sugratrans}).

Let's begin with the metric. First of all, the spatial vector $\delta_\zeta g_{0i}$ must vanish. Using (\ref{sugratetrad}) and (\ref{ans}), we get
\begin{align}\label{vanishvector}
	\delta_\zeta g_{0i}=i (\psi_0 \sig_i \bze-\zeta \sig_i \bar{\psi}_0)-i (A_1 \psi_0 \sig_i \bze-A_1^* \zeta \sig_i \bar{\psi}_0)
\end{align}
This fixes the constant $A_1=1$. The off-diagonal spatial components yield no further restriction.  

Having been left with a single independent spinor $\psi_0$, we write it as $\psi_{0 \alpha}=-\frac{1}{2} \sigma_{0 \al \dal} \bph^\dal$, which leads to the covariant expression\footnote{In conformal supergravity, the $S$-supersymmetry transformations can be used to gauge away $\bar{\phi}$. However such transformations are excluded in Einstein's supergravity \cite{kuzenko}.}
\begin{align}\label{ansatz}
	\psi_m^{\ \alpha}=\frac{1}{2} \bar{\phi}_{\dot{\alpha}} \bar{\sigma}_m^{\dot{\alpha} \alpha}.
\end{align}
Thus, imposing spherical symmetry on $\psi_m \psi_n$ removes the traceless part of the Rarita-Schwinger field (cf. (\ref{decomposed})), just like it removes the spatial vector $g_{0i}$ and the traceless symmetric part of the spatial metric $h_{\langle ij\rangle}$ \cite{ellis}.

This form of the vector-spinor, related to conformal flatness of the corresponding superspace \cite{kuzenko}, seems appropriate given that FLRW spacetimes are conformally flat \cite{ibison,lihoshi}.

Further, (\ref{ansatz}) yields
\begin{align}
\psi_m \psi_n=\frac{1}{4} g_{mn} \bar{\phi}_\dal \bar{\phi}^\dal
\end{align}
which is of the form (\ref{gfrw}) provided that
\begin{align}\label{invariantsquare}
\delta_X \bph \bph\equiv -X^l \p_l \big[\bar{\phi}_\dal \bar{\phi}^\dal\big]=0,
\end{align}
where $X^l$ denotes any of the k=0 Killing vectors. Therefore, $\bph \bph$ must depend only on time, but $\bph$ may depend on space through a local Lorentz transformation.

So far, we have used the off-diagonal components of (\ref{preserve}) to refine the form of the Rarita-Schwinger field implied by (\ref{gfrw}):  By giving up the function $A_1$, we avoid imposing a constraint on $\zeta$. We still have to consider the diagonal components.

For the vector-spinor (\ref{ansatz}), the supergravity transformation of the tetrad (\ref{sugratetrad}) takes the form
\begin{align}\label{conformal}
	\delta_\zeta e_m^{\ a}&=\frac{i}{2} e_{ma} (\bar{\phi} \bsig^b \sig^a \bar{\zeta}-\zeta \sig^a \bsig^b \phi) \nonumber \\
	&=e_{mb} M^{ba}+\Omega e_m^{\ a},
\end{align}
with the field-dependent Lorentz generator and Weyl factor given, respectively, by
\begin{subequations}
\begin{align}
	M^{ba}&=i (\bar{\phi} \bar{\sigma}^{ba} \bar{\zeta}-\zeta \sigma^{ab} \phi), \label{spingen} \\
	\Omega&=\frac{i}{2} (\zeta \phi-\bar{\phi} \bar{\zeta}). \label{cfactor}
\end{align}
\end{subequations}
Thus, the infinitesimal supergravity transformation of the metric is
\begin{equation}\label{confor}
	\delta_\zeta g_{mn}=2\Omega g_{mn}=i (\zeta \phi-\bar{\phi} \bar{\zeta}) g_{mn}.
\end{equation}
with $g_{mn}$ the k=0 FLRW metric. Thus, (\ref{preserve}) holds provided that $\Omega$ is a function of time only: $\delta_X \Omega=-X^l \p_l \Omega=0$. To avoid complication dealing with real and imaginary parts separately, we simply impose that the whole complex quantity depends only on time,
\begin{align}\label{invsupergrav}
	\delta_X [\zeta \phi]\equiv -X^l \p_l [\zeta^\al \phi_\al]=0.
\end{align}
This first constraint follows from (\ref{preserve}).

Now, we continue with (\ref{preserve2}). For simplicity, we omit auxiliary fields since they vanish anyway in the vacuum case. 

For (\ref{ansatz}), the contorsion (\ref{contorsion}) is given by
\begin{align}\label{contor}
	\kappa_{m\beta \alpha}=\frac{i}{16} \big[\sig_{m\be \dga} \bar{\phi}^\dga \phi_\alpha+\sigma_{m\al \dga} \bar{\phi}^\dga \phi_\beta\big].
\end{align}
Thus the supergravity transformation rule (\ref{sugraspinor}) yields\footnote{We use the rearrangement $\chi_\al \xi_\be=\frac{2-d}{2} \chi_\al \xi_\be+\frac{d}{2} \chi_\be \xi_\al+\frac{d}{2} \epsilon_{\alpha \beta} \chi^\gamma \xi_\gamma$, with $d$ an arbitrary constant.}
\begin{align}\label{deltaspin}
	\delta_\zeta \psi_m^{\ \al}&=-2 D_m \zeta^\al-\frac{3 i}{16} \zeta \phi (\bph \bar{\sigma}_m)^\al+\frac{1}{8} \bph_\dde \bsig_m^{\dde \de} M_\de^{\ \al}
\end{align}
where $D_m$ is the covariant derivative with bosonic connection (see (\ref{covz})), and the spinor form of  (\ref{spingen}) is 
\begin{align}\label{eme}
	M_{\beta \alpha}=\frac{i}{2} [\zeta_\beta \phi_\alpha
	+\zeta_\alpha \phi_\beta].
\end{align}

Now, contracting (\ref{deltaspin}) with $\psi_{n\alpha}$ and symmetrizing in $m, n$, we obtain
\begin{align}\label{deltaspinmet}
	\delta_\zeta [\psi_m \psi_n]=(D_m \zeta) \sig_n \bar{\phi}+(D_n \zeta) \sig_m \bar{\phi}-\frac{3 i}{16} \zeta \phi\, g_{mn} \bph \bph.
\end{align}
Therefore, the remaining condition to satisfy (\ref{preserve2}) is
\begin{align}\label{condit}
	\delta_X  V_{mn} \equiv \delta_X [(D_m \zeta) \sig_n \bar{\phi}+(D_n \zeta) \sig_m \bar{\phi}]=0.
\end{align}
This latter condition yields further restrictions on the transformation parameter as shown below.

\subsubsection{k=0 FLRW}\label{flat}
In Cartesian coordinates $(x^0,x^1,x^2,x^3)=(t,x,y,z)$ the metric is $ds^2=-N^2 dt^2+a^2 (dx^2+dy^2+dz^2)$. For the diagonal tetrad $\bm e^{\tilde{0}}=-N dt$, $\bm e^{\tilde{k}}=dx^k \delta_{k}^{\ \tilde{k}} a$ (tildes denote Lorentz values, see Appendix \ref{appendix1}), we obtain the following connection components
\begin{align}\label{conflat}
	\omega_{i \beta}^{\ \ \alpha}=\frac{\dot{a}}{2a} (\sigma_i \bsig^0)_\be^{\ \al}, && \omega_{i\ \dbe}^{\ \dal}=\frac{\dot{a}}{2 a} (\bsig_i \sigma^0)^\dal_{\ \dbe}.
\end{align}
Assuming $\zeta$ and $\bph$ depend only on time, the bosonic covariant derivatives are
\begin{align}\label{derivs}
	D_0 \zeta^\al=\p_t \zeta^\al, && D_i \zeta^\al=\frac{\dot{a}}{2a} (\zeta \sigma_i \bsig^0)^\al
\end{align}
Now, in analogy with (\ref{confor}), we satisfy (\ref{condit}) by taking $V_{mn}\propto g_{mn}$. Using (\ref{derivs}), we get
\begin{subequations}
\begin{align}
	V_{i0}&=\Big(\dot{\zeta}-\frac{\dot{a}}{2a} \zeta\Big) \sig_i \bph, \\
	V_{ij}&=\frac{\dot{a}}{a} g_{ij} \zeta \sigma^0 \bph \propto \delta_{ij}. \label{vij}
\end{align}
\end{subequations}
Thus we impose the following restriction on the transformation parameter
\begin{align}\label{conforzeta}
	\dot{\zeta}^\al-\frac{\dot{a}}{ 2a} \zeta^\al=0.
\end{align}
where $\dot{a}=da/dt$. Then equations (\ref{condit}) hold since 
\begin{align}
V_{mn}=\frac{\dot{a}}{N a} (\zeta \sig_{\tilde{0}} \bph) g_{mn}.
\end{align}
In other system of coordinates or frames of reference, $\zeta$ and $\bph$ may depend on the spatial coordinates; this freedom is utilized in the solution worked out using spherical coordinates detailed in Appendix \ref{sphere}. 

\subsubsection{k=1 FLRW}
In comoving coordinates $(t,\chi,\theta,\varphi)$, the metric is given by $ds^2=-N^2 dt^2+a^2 [d\chi^2+\sin^2 \chi d\theta^2+\sin^2 \chi \sin^2 \theta d\varphi^2]$ \cite{inverno,ellis}. For the diagonal tetrad 
\begin{align}
e_m^{\ a}=\begin{bmatrix}
		-N & 0 & 0 & 0 \\
		0 & a & 0 & 0 \\
		0 & 0 & a \sin \chi & 0 \\
		0 & 0 & 0 & \sin \chi \sin \theta \\
	\end{bmatrix}
\end{align}
the non-vanishing connection components are
\begin{subequations}\label{kone}
	\begin{align}
		\omega_{\chi \beta}^{\ \ \ \al}&=\frac{\dot{a}}{2N} (\sig^{\tilde{0}} \bsig^{\tilde{1}})_\beta^{\ \al}, \\
		\omega_{\theta \beta}^{\ \ \ \al}&=\frac{\dot{a}}{2N} \sin \chi (\sig^{\tilde{0}} \bsig^{\tilde{2}})_\beta^{\ \al}+\frac{i}{2} \cos \chi (\sig^{\tilde{0}} \bsig^{\tilde{3}})_\beta^{\ \al}, \\
		\omega_{\varphi \beta}^{\ \ \ \al}&=\frac{\dot{a}}{2N} \sin \chi \sin \theta (\sig^{\tilde{0}} \bsig^{\tilde{3}})_\beta^{\ \al}-\frac{i}{2} \cos \chi \sin \theta (\sig^{\tilde{0}} \bsig^{\tilde{2}})_\beta^{\ \al}+\frac{i}{2} \cos \theta (\sig^{\tilde{0}} \bsig^{\tilde{1}})_\beta^{\ \al}.
	\end{align}
\end{subequations}

The vanishing of the time-space components $V_{i0}$ (\ref{condit}) yields the set of equations
\begin{subequations}\label{sph}
\begin{align}
0=&\frac{1}{N} \Big(\frac{\dot{a}}{2 a} \zeta-\partial_t \zeta\Big) \sigma^{\tilde{1}} \bph-\frac{1}{a} (\partial_\chi \zeta) \sigma^{\tilde{0}} \bph, \\
0=&\frac{1}{N} \Big(\frac{\dot{a}}{2 a} \zeta-\partial_t \zeta\Big) \sigma^{\tilde{2}} \bph+\frac{1}{a \sin \chi} \Big(\frac{\cos \chi}{2} \zeta \sigma^{\tilde{1}} \bsig^{\tilde{2}}-\partial_\theta \zeta\Big) \sigma^{\tilde{0}} \bph, \\
0=&\frac{1}{N} \Big(\frac{\dot{a}}{2 a} \zeta-\partial_t \zeta\Big) \sigma^{\tilde{3}} \bph+\frac{1}{a \sin \chi \sin \theta} \Big(\frac{\cos \chi \sin \theta}{2} \zeta \sigma^{\tilde{1}} \bsig^{\tilde{3}}+\frac{\cos \theta}{2} \zeta \sigma^{\tilde{2}} \bsig^{\tilde{3}}-\partial_\varphi \zeta\Big) \sigma^{\tilde{0}} \bph.
	\end{align}
\end{subequations}

In this case, we consider a modification of (\ref{conforzeta}), namely,
\begin{align}\label{confor2}
	\frac{\partial_t \zeta}{N}+\Big(\frac{u}{a}-\frac{\dot{a}}{2 Na}\Big) \zeta =0,
\end{align}
where $u$ is a complex constant, then (\ref{sph}) becomes
\begin{subequations}\label{spatder}
	\begin{align}
		0=&u \zeta \sigma^{\tilde{1}} \bar{\sigma}^{\tilde{0}}-\partial_\chi \zeta, \\
		0=&u \sin \chi \zeta \sigma^{\tilde{2}} \bar{\sigma}^{\tilde{0}}+\frac{i}{2} \cos \chi \zeta \sigma^{\tilde{3}} \bar{\sigma}^{\tilde{0}}-\partial_\theta \zeta \\
		0=&u \sin \chi \sin \theta \zeta \sigma^{\tilde{3}} \bar{\sigma}^{\tilde{0}}-\frac{i}{2} \cos \chi \sin \theta \zeta \sigma^{\tilde{2}} \bar{\sigma}^{\tilde{0}}+\frac{i}{2} \cos \theta \zeta \sigma^{\tilde{1}} \bar{\sigma}^{\tilde{0}}-\partial_\varphi \zeta
	\end{align}
\end{subequations}

Now, proceeding as in the example of Appendix \ref{sphere}, we obtain a solution of the form 
\begin{align}
\zeta=\tilde{\zeta}^\ga(t) \Lambda_\ga^{\ \alpha}(\chi,\theta,\varphi)
\end{align}
where $\tilde{\zeta}$ satisfies (\ref{confor2}) and, setting $u=i/2$, the (finite) Lorentz transformation
\begin{subequations}
 \begin{align}
	\Lambda_1^{\ 1}&=\frac{1}{\sqrt{2}} e^{i \varphi/2} \big(\cos \frac{\theta+\chi}{2}-i \sin \frac{\theta-\chi}{2}\big) \\
	\Lambda_1^{\ 2}&=-\frac{1}{\sqrt{2}} e^{i \varphi/2} \big(\cos \frac{\theta-\chi}{2}+i \sin \frac{\theta+\chi}{2}\big) \\
	\Lambda_2^{\ 1}&=e^{-i \varphi/2} \big(\cos \frac{\theta-\chi}{2}-i \sin \frac{\theta+\chi}{2}\big) \\
	\Lambda_2^{\ 2}&=e^{-i \varphi/2} \big(\cos \frac{\theta+\chi}{2}+i \sin \frac{\theta-\chi}{2}\big).
\end{align}
\end{subequations}

Using (\ref{kone}) and (\ref{spatder}), we obtain
\begin{align}
	V_{mn}&=\Big(\frac{\dot{a}}{Na}-\frac{i}{a} \Big) (\zeta \sigma_{\tilde{0}} \bph) g_{mn}.
\end{align}
Finally, to satisfy (\ref{invsupergrav}), we must choose $\phi_\al=\Lambda_\al^{\ \be} \tilde{\phi}_\be(t)$.

Equations (\ref{confor2}) and (\ref{spatder}) constitute a generalization to time dependent radius (with vanishing auxiliary fields) of the supersymmetric background $\mathbb{R}\times S^3$ \cite{Festuccia}, although here we are not considering matter yet.

So far, we have used the softened Killing equations (\ref{quadratic1}) to reduce the general supergravity multiplet to the lapse, scale factor and a single Weyl spinor. Further, the preservation of the quadratic equations under supergravity yielded constraints such as (\ref{invsupergrav}) and (\ref{confor}) for the spatially flat case. 

\subsection{FLRW multiplet}\label{supermiltiplet}
The vector-spinor can be expressed in terms of its trace and traceless components as follows \cite{wessbagger}
\begin{align}\label{decomposed}
	\psi_m^{\ \alpha}=\frac{1}{2} \bar{\phi}_{\dga} \bar{\sigma}_m^{\dga \alpha}-\frac{1}{2} \bar{\sigma}_m^{\dga \ga} W_{\ga \ \dga}^{\ \al}.
\end{align}
where 
\begin{align}
	\bar{\phi}_\dal &\equiv -\frac{1}{2} \psi_n^{\ \al} \sigma^n_{\al \dal}, \label{def}\\
	W_{\be \alpha \dga} &\equiv \tf \big(\sigma^n_{\be \dga} \psi_{n \al}+\sigma^n_{\al \dga} \psi_{n \be}\big) \label{traceless},
\end{align}
where $W_n^{\ \al} \sig^n_{\al \dal}=-\tf W_{\be \ \dga}^{\ \al} \bsig_n^{\be \dga} \sig^n_{\al \dal}=0$. The tetrad also enters (\ref{decomposed}) through the $\sigma$-matrices.

Therefore, the solution of the quadratic Killing equations (\ref{quadratic1}) sets (\ref{traceless}) to zero. Just like generic supergravity transformations do not preserve $\psi_m=0$ automatically, they do not preserve $W_{\be \al \dga}=0$. Thus, according to our second guiding principle, we additionally require that,
\begin{align}\label{conformalkilling}
	\delta_\zeta W_{\be \al \dga}=0.
\end{align}

Using definition (\ref{traceless}), together with (\ref{conformal}) and (\ref{deltaspin}), we obtain
\begin{align}
	\delta_\zeta W_{\be \al \dga}&=\delta_\zeta \psi_{n (\be} \sigma^n_{\al) \dga}+\psi_{n(\be} \sigma^a_{\al) \dga} \delta_\zeta e_a^{\ n} \nonumber \\
	&=-\sig^n_{\be \dga} D_n \zeta_\al-\sig^n_{\al \dga} D_n \zeta_\be+\frac{3}{4}\bar{\phi}_\dga M_{\be \al}. \label{vanishconf}
\end{align}
In the k=0 FLRW case, using (\ref{derivs}) and (\ref{conforzeta}), we have
\begin{align}
	\sig^n_{\be \dga} D_n \zeta_\al&=\sig^0_{\be \dga} D_0 \zeta_\al+\sig^k_{\be \dga} D_k \zeta_\al
 \nonumber \\	&=\frac{\dot{a}}{2a} \Big(\sig^0_{\be \dga} \zeta_\al+\sig^k_{\be \dga} \sig^0_{\al \dde} \bsig_k^{\dde \de} \zeta_\de\Big) \nonumber \\
	%&=\frac{\dot{a}}{2a} \Big(\sig^0_{\be \dga} \zeta_\al+\sig^n_{\be \dga} \sig^0_{\al \dde} \bsig_n^{\dde \de} \zeta_\de-\sig^0_{\be \dga} \sig^0_{\al \dde} \bsig_0^{\dde \de} \zeta_\de \Big) \nonumber \\
%	&=\frac{\dot{a}}{2a} \Big(2 \sig^0_{\be \dga} \zeta_\al-2 \sig^0_{\al \dga} \zeta_\be\Big)
&=\frac{\dot{a}}{a} \ep_{\be \al} \zeta^\ga \sig^0_{\ga \dga}. 
\end{align}
Thus, it remains in (\ref{vanishconf}) the term due to the field-dependent Lorentz transformation (\ref{spingen}), which cannot be removed nor with a coordinate nor Lorentz transformation since $W_{\be \al \dga}$ is a vanishing Lorentz tensor. Therefore, we must impose a further constraint $M_{\be \al}=0$ or, equivalently, 
\begin{align}\label{lorcons}
	\zeta_\be \phi_\al=\tf \ep_{\be \al} \zeta \phi, && \bze_\dbe \bph_\dal=-\tf \ep_{\dbe \dal} \bze \bph,
\end{align}
that still admit nontrivial solution.

Under this extra condition, using definition (\ref{def}), the transformation of the spin-$\tf$ component is
\begin{align}\label{deltaphi}
	\delta_\zeta \bar{\phi}_\dal&=-\tf (\delta_\zeta \psi_n^{\ \ga}) \sigma^n_{\ga \dal}-\frac{1}{2} \psi_n^{\ \ga} \sigma^a_{\ga \dot{\alpha}} \delta_\zeta e_a^{\ n} \nonumber \\
	&=\sigma^k_{\al \dal} D_k \zeta^\alpha-\frac{3 i}{8} \zeta \phi \bph_\dal-\frac{i}{2} (\zeta \phi-\bar{\zeta} \bph) \bph_\dal.
\end{align}

From (\ref{confor}), the scale factor and lapse transform as 
\begin{subequations}\label{scale}
\begin{align}
	\delta_\zeta N&=\frac{i}{2} N (\zeta \phi-\bar{\phi} \bar{\zeta}), \label{time} \\
\delta_\zeta a&=\frac{i}{2} a (\zeta \phi-\bar{\phi} \bar{\zeta}), \label{scafac} \\
\delta_\zeta \bar{\phi}_\dal&=2 \frac{\dot{a}}{a} \zeta^\al \sig^0_{\al \dal}-\frac{7 i}{8} \zeta \phi \bph_\dal.
\end{align}
\end{subequations}
Note that (\ref{time}) is only consistent with (\ref{scafac}) for the gauge choice $N=a$ corresponding to conformal time ($N=1$ is excluded since we choose $\psi_0 \ne 0$ when solving (\ref{sec:2})).

Now, the general commutator of supergravity transformations yields a gauged translation with parameter
\begin{align}\label{parameter}
	\xi^a =2i \big[\bar{\eta} \bsig^a \zeta-\bar{\zeta} \bsig^a \eta\big],
\end{align}

For the vector-spinor (\ref{ansatz}), we obtain 
\begin{subequations}\label{coma}
\begin{align}
[\delta_\eta, \delta_\zeta] N&=-\p_t [N \xi^0]=2i \p_t [N (\bze \bsig^0 \eta-\bet \bsig^0 \zeta)], \\
[\delta_\eta, \delta_\zeta] a&=-\xi^0 \dot{a}=2i [\bze \bsig^0 \eta-\bet \bsig^0 \zeta] \dot{a}.
\end{align}
\end{subequations}

On the other hand, using (\ref{def}) we get 
\begin{align}\label{com}
	[\delta_\eta, \delta_\zeta] \bar{\phi}_{\dot{\alpha}}=-\xi^m D_m \bph_\dal+D_m \xi^\al \sigma^m_{\al \dal}+\frac{9i}{16} \bar{\phi} \bar{\phi} \bar{\xi}_\dal, 
\end{align}
where $\xi^\alpha=i (\bph \bet \zeta^\al-\bph \bze \eta^\al)$. Expanding this in the k=0 FLRW case, using (\ref{conflat}), (\ref{conforzeta}) and (\ref{lorcons}), (\ref{com}) reduces to
\begin{align}
	[\delta_\eta, \delta_\zeta] \bar{\phi}_{\dot{\alpha}}&=2i \big[\dot{\bph} \bze \eta^\al-\dot{\bph} \bet \zeta^\al+2 \frac{\dot{a}}{a} (\bph \bet \zeta^\al-\bph \bze \eta^\al) \big]\sig^0_{\al \dal}.
\end{align}

We have exhausted all the requirements by imposing restrictions (\ref{confor}) or (\ref{confor2}), and (\ref{lorcons}). The remaining transformation parameters satisfy strong restrictions that, nonetheless admits nontrivial solutions by using a sufficiently large basis of the Grassmann algebra to which the dynamical variables and transformation parameters belong \cite{Finkelstein}. These additional contraints do not conflict with the W-Z gauge fixing since the left transformation parameters are completely arbitrary until they are fixed by a particular solution.

As mentioned in Section \ref{intro}, our approach to FLRW supergravity can be seen as a relaxed spinor Killing equation in the sense that, instead of $\delta_\zeta e_m^{\ a}=0$ and $\delta_\zeta \psi_m^{\ \al}=0$, we have
\begin{subequations}
	\begin{align}
		&\delta_\zeta e_m^{\ a}=\Omega e_m^{\ a}, \\
		&\delta_\zeta \psi_m \to  
		\begin{cases}
			\delta_\zeta \phi^\al \ne 0 \\
			\delta_\zeta W_{\be \al \dga}=0
		\end{cases}
	\end{align}
\end{subequations}
with $\Omega$ given in (\ref{cfactor}).

\section{Generalization to arbitrary time gauge}\label{sec:3}
In the general theory, lapse $N$, shift-vector $N_i$ \cite{bojowald} and $\psi_0$ are all gauge fields that can be specified in an arbitrary way. Keeping this freedom available is important in the canonical formulation where these fields appear as Lagrange multipliers of the first-class constraints \cite{deathbook,moniz}. In order to leave $N$ and $\psi_0$ arbitrary we have to allow non-vanishing spatial vectors such as $N_i$ and $\psi_0 \psi_i$. 

In this sense, perhaps a more appropriate definition of generalized isometries in supergravity should include configurations that, in a certain gauge, satisfy the quadratic equations (\ref{quadratic1}). For the spatial isometries that concern us in this work, this relaxation allows the tetrad and vector-spinor used in the construction of the FLRW supersymmetric model in \cite{death88,death92}, namely,
\begin{align}\label{tetradfrw}
	e_m^{\ a}&=\begin{bmatrix}
		-N & a N^k E_k^{\ A} \\
		0 & a E_i^{\ A} 
	\end{bmatrix}, && 
	e_a^{\ m}=\begin{bmatrix}
		-N^{-1} & N^{-1} N^j \\
		0 & a^{-1} E_A^{\  j}
	\end{bmatrix},
\end{align}
($A, B, C$ denote spatial Lorentz indices) and 
\begin{align}\label{spinproj}
	\psi_0^{\ \al}=\psi_0^{\ \al}(t), && \psi_i^{\ \al}=\frac{2}{3} \bet_\dga(t) \bsig_i^{\dga \al} 
\end{align}
with the spatial tetrad and vector-spinor satisfying respectively,
\begin{align}\label{espacial}
	\delta_X (E_i^{\ A} E_{j A})=0, && \delta_X (\psi_i^{\ \al} \psi_{j \al})=0.
\end{align} 
Upon fixing the gauge $N_i=0$ and $\psi_0=\frac{2}{3} \bet_\dga(t) \bsig_0^{\dga \al}$, equations (\ref{quadratic1}) hold.

\subsection{Compatible supergravity transformations}
Since the gauge fields $N, N_i$ are accommodated in $e_0^{\ a}$ and $\psi_0$ only, the remaining consideration of supergravity transformations will focus on the spatial tetrad and vector-spinor.

Using the supergravity transformation law of the tetrad (\ref{sugratetrad}) and (\ref{spinproj}), we have
\begin{align}\label{contetrad}
	\delta_\zeta e_i^{\ a}=e_{ib} M^{ba}+e_i^{\ a} \Omega
\end{align}
where 
\begin{subequations}
\begin{align}\label{newm}
	M^{ba}&=\frac{4i }{3} (\bet \bsig^{ba} \bar{\zeta}-\zeta \sigma^{a b} \eta), \\
	\Omega&=\frac{2 i}{3} (\zeta \eta-\bet \bar{\zeta}).
\end{align}
\end{subequations}

Now, for a tetrad (\ref{tetradfrw}) (with $N_i=0$) and spinor-vector of the form (\ref{spinproj}), the contorsion components (\ref{contorsion}) are given by 
\begin{align}\label{spatcon}
	\kappa_{i\beta \alpha}=-\frac{i}{12} (\psi_0 \eta-\bet \bar{\psi}_0) \big(\ep_{\al \ga} \sigma^0_{\beta \dga} \bar{\sigma}_i^{\dga \ga}+\ep_{\be \ga} \sig^0_{\al \dga} \bsig_i^{\dga \ga}\big)+\frac{i}{9} \big(\sig_{i \be \dot{\delta}} \bet^{\dot{\delta}} \eta_\al+\bsig_{i \al \dot{\delta}} \bet^{\dot{\delta}} \eta_\beta\big) 
\end{align}

Next, the supergravity transformation of the spatial spinor-vector reads, on-shell,
\begin{align}\label{sh0}
	\delta_\zeta \psi_i^{\ \alpha}=-2 D_i \zeta^\alpha+\frac{i}{3} (\psi_0 \eta-\bet \bar{\psi}_0) (\zeta \sigma^0 \bsig_i)^\alpha-\frac{i}{3} \zeta \eta (\bet \bsig_i)^\alpha+\frac{1}{6} \bet_\dde \bsig_i^{\dde \be} M_\be^{\ \al}.
\end{align}
where
\begin{align}
	M_{\be \al}=\frac{2 i}{3} (\zeta_\be \eta_\al+\zeta_\al \eta_\be)
\end{align}

Contracting (\ref{sh0}) with $\psi_{j\al}$ and symmetrizing in $i,j$, we get
\begin{align}\label{psipsi}
	\delta_\zeta (\psi_i \psi_j)=\frac{4}{3} \big(D_i \zeta \sig_j \bet+D_j \zeta \sig_i \bet\big)-\frac{4 i}{9} \zeta \eta g_{ij} \bet \bet+\frac{4 i}{9} (\psi_0 \eta-\bet \bar{\psi}_0) \zeta \sigma^0 \bet g_{ij}.
\end{align}
which satisfies (\ref{espacial}) (cf. (\ref{vij})).

Unlike the covariant case, here the transformation parameter $\zeta(t)$ is no longer restricted by (\ref{conforzeta}) or (\ref{confor2}). 

\subsection{FLRW multiplet with arbitrary time gauge}
As in the covariant decomposition, the spatial spinor-vector is decomposed as
\begin{align}\label{spindrop}
	\psi_i^{\ \al}=\frac{2}{3} \bar{\eta}_\dga \bsig_i^{\dga \al}-\tf \Upsilon_{\ga \ \dga}^{\ \al} \bsig_i^{\dga \ga}.
\end{align}
where we use the following definitions \cite{moniz}
\begin{subequations}
\begin{align}
	\bet_\dal&\equiv -\tf \psi_k^{\ \al} \sigma^k_{\al \dal}, \label{trace} \\
	\Upsilon_{\be \al \dga}&=\psi_{k (\be} \sig^k_{\al) \dga}+\frac{1}{3} \big[\sig^{\tilde{0}}_{\be \dga} \sig_{\tilde{0} \al \dde}+\sig^{\tilde{0}}_{\al \dga} \sig_{\tilde{0} \be \dde}\big] \bet^\dde \label{purespin}
\end{align}
\end{subequations}
such that
\begin{align}
	\Upsilon_k^{\ \al} \sig^k_{\al \dbe}=-\tf \Upsilon_{\be \ \dga}^{\ \al} \bsig_k^{\dga \be} \sig^k_{\al \dbe}=0. 
\end{align}

As in the previous section, we impose the constraint $M_{\be \al}=0$ or
\begin{align}\label{vanishbroken}
\zeta_\beta \eta_\al=\tf \ep_{\be \al} \zeta \eta.
\end{align}

Now, $\delta_\zeta \Upsilon_{\be \al \dga}$ follows from definition (\ref{purespin}). Under (\ref{vanishbroken}), $\delta_\zeta e_i^{\ a}=\Omega  e_i^{\ a}$ and the contribution to  due to the inverse tetrad in (\ref{purespin}) yields a term proportional to $-\Omega \Upsilon_{\be \al \dga}=0$. The remaining contribution comes from the transformation of the vector-spinor. Using (\ref{sh0}), we have
\begin{align}\label{deltatilde}
	\delta_\zeta \Upsilon_{\be \al \dga}=&-\sig^k_{\be \dga} \D_k \zeta_\al-\sig^k_{\al \dga} \D_k \zeta_\be+\frac{1}{3} (\sig^{\tilde{0}}_{\be \dga} \sig_{{\tilde{0}} \al \dde}+\sig^{\tilde{0}}_{\al \dga} \sig_{{\tilde{0}} \be \dde})
	\sigma^{k\dde \ga} \D_k \zeta_\ga.
\end{align}
By virtue of (\ref{vanishbroken}), the covariant derivative is of the form $\D_i \zeta_\al=\sig_{i \al \dbe} \gamma^{\dbe}$, as can be seen in (\ref{sh0}), consequently (\ref{deltatilde}) vanishes.

On the other hand, from definition (\ref{trace}),
\begin{align}
	\delta_\zeta \bet_\dal &\equiv -\tf \sigma^k_{\al \dal} \delta_\zeta \psi_k^{\ \al}-\tf \psi_k^{\ \al} \sigma^b_{\al \dal} \delta_\zeta e_b^{\ k}.
\end{align}
Using (\ref{contetrad}) and (\ref{sh0}) subject to (\ref{vanishbroken}), we obtain
\begin{align}\label{deltatrace}
	\delta_\zeta \bet_\dal=&\sigma^k_{\al \dal} D_k \zeta^\alpha+\frac{i}{2} (\psi_0 \eta-\bet \bar{\psi}_0) (\zeta \sigma^0)_\dal-\frac{7 i}{6} \zeta \eta \bet_\dal.
\end{align}

Thus, in the k=0 FLRW this scenario, we have
\begin{align}
	\delta_\zeta N&=-\frac{i}{N} (\psi_0 \sigma_0 \bar{\zeta}-\zeta \sigma_0 \bar{\psi}_0), \\
	\delta_\zeta a&=\frac{2 i}{3} a (\zeta \eta-\bet \bar{\zeta}), \\
	\delta_\zeta \bet_\dal&=\Big(\frac{3 \dot{a}}{2a}+\frac{i}{2} (\psi_0 \eta-\bet \bar{\psi}_0)\Big) (\zeta \sigma^0)_\dal-\frac{7i}{6} \zeta \eta \bet_\dal, \label{newspin} \\
	\delta_\zeta \psi_0^{\ \al}&=-2 \partial_t \zeta^\alpha-\frac{2 i}{3} \zeta \eta \psi_0^{\ \al}-\frac{i}{3} (\psi_0 \eta) \zeta^\al+\frac{2}{9} \zeta \sigma_0 \bet \ \eta^\al+\frac{2}{9} \zeta \eta (\bet \bsig_0)^\al.
\end{align}
using $\omega_{0\be}^{\ \ \al}=0$ and the contorsion component
\begin{align}
	\kappa_{0\beta \alpha}=-\frac{i}{6} (\psi_{0\beta} \eta_\al+\psi_{0 \al} \eta_\be)-\frac{1}{9} (\sigma_{0 \beta \dga} \bet^\dga \eta_\al+\sigma_{0 \al \dga} \bet^\dga \eta_\be).
\end{align}

Setting $\psi_0=\tf \bph_\dal \bsig_0^{\dal \al}$ and $\bet_\dal=\frac{3}{4} \bph_\dal$, we recover the transformation rules (\ref{scale}). If we further restrict $\zeta(t)$ by (\ref{conforzeta}), supergravity transformation will preserve the FLRW form (\ref{gfrw}).

\subsubsection{Generalization to complex 4-geometry }
Following \cite{death88,death92}, we consider the supergravity transformation followed by the field-dependent Lorentz transformation $M_{\be \al}$ (\ref{newm}), which for the moment is not constrained to vanish. Then we have
\begin{align}\label{newtetrad}
	\delta_\zeta^* e_i^{\ a} \equiv (\delta_\zeta-\delta_M)  e_i^{\ a}=\frac{2 i}{3} e_i^{\ a} (\zeta \eta-\bet \bar{\zeta})
\end{align}

On the other hand, the transformation of the vector-spinor, can be arranged as 
\begin{align}
	\delta_\zeta^* \psi_i^{\ \alpha} & \equiv (\delta_\zeta-\delta_M) \psi_i^{\ \alpha} \nonumber \\
	&=-2 D_i \zeta^\alpha+\frac{i}{3} (\psi_0 \eta-\bet \bar{\psi}_0) (\zeta \sigma^0 \bsig_i)^\alpha-\frac{(2+d) i}{9} \zeta \eta (\bet \bsig_i)^\alpha+\frac{(2-d) i}{9} (\bet \bsig_i \zeta) \eta^\alpha \nonumber \\
	&-\frac{d i}{9} (\bet \bsig_i \eta) \zeta^\al+\frac{4 i}{9} (\bet \bsig_i \eta) \zeta^\al-\frac{4 i}{9}  (\bet \bsig_i \zeta) \eta^\al 
\end{align}
where $d$ is an arbitrary constant. Choosing  $d=-2$, we obtain
\begin{align}\label{terra2}
	(\delta_\zeta-\delta_M) \psi_i^{\ \alpha}=-2 D_i \zeta^\alpha+\frac{i}{3} (\psi_0 \eta-\bet \bar{\psi}_0) (\zeta \sigma^0 \bsig_i)^\alpha+\frac{2 i}{3} (\bet \bsig_i \eta) \zeta^\al
\end{align}

Now, for $\delta_\zeta^* \Upsilon_{\be \al \dga}=0$, we must impose a different constraint 
\begin{align}\label{onedl}
	\bet \bsig_i \eta=0,
\end{align}
which is interpreted as the 1D version of the Lorentz constraint of canonical supergravity \cite{deathbook,moniz} since it implies
\begin{align}\label{lorentz1d}
	J_{\be \al}\equiv \eta_\be \sig^{\tilde{0}}_{\al \dga} \bet^\dga+\eta_\al \sig^{\tilde{0}}_{\be \dga} \bet^\dga=0. %, && 	\tilde{J}_{\dal \dbe}\equiv \eta^\de \sig^0_{\de \dal} \bph_\dbe+\eta^\de \sig^0_{\de \dbe} \bph_\dal=0.
\end{align}
A nontrivial solution of which requires that $\bar{\eta}\ne \eta^*$.  

The transformation law of the spin-$\tf$ (\ref{newspin}) changes to
\begin{align}\label{final}
\delta_\zeta^* \bet_\dal&=\Big(\frac{3 \dot{a}}{2a}+\frac{i}{2} (\psi_0 \eta-\bet \bar{\psi}_0)\Big) (\zeta \sigma^0)_\dal-\Omega \bet_\dal.
\end{align}

For (\ref{onedl}) to be preserved under (\ref{final}) throws the constraint (\ref{vanishbroken}). Nonetheless, in the canonical formulation it holds by virtue of the supersymmetric constraints \cite{death92}.

\section{Conclusions}\label{sec:4}
We explored some FLRW reductions of 4D N=1 supergravity (in the Wess-Zumino gauge) arising from imposing the FLRW isometries on the quadratic form $\psi_m \psi_n$ instead of the vector-spinor alone. This form is motivated by a Lorentz contraction of vielbein superfields $E_m^{\ A} E_{mA}$.

Two guiding principles are used in our derivation. The first consists of finding a working solution of the  quadratic Killing equations (\ref{quadratic1}), and identifying the subset of supergravity transformations that preserve the Killing equations via (\ref{superpres}). Second, ensure that the leftover components form a supermultiplet. For the FLRW example, the second criteria translates into the extra condition (\ref{conformalkilling}). The solution for the vector-spinor (\ref{ansatz}) is expressed in terms of its spin-$\frac{1}{2}$ component, $\phi^\al$, whereas its spin-$\frac{3}{2}$ component, $W_{\be \al \dga}$, vanishes. This makes sense since the FLRW metric does not possess spin-2 component.

The homogeneous spinor $\phi$ is not related to the gravitino fluctuation of phenomenological supergravity. Rather, it is part of our background FLRW supermultiplet. In fact, the supergravity transformation of $a(t)$ is linear in $\phi$. A fermion partner of the scale factor is a hallmark of supersymmetric cosmology \cite{death92,moniz,ramirez}.

From the set of general supergravity transformations allowed by the W-Z gauge, those that comply with our two principles are defined by constraints (\ref{invsupergrav}), (\ref{conforzeta}) and (\ref{lorcons}), for the k=0 case (for k=1 (\ref{confor2} replaces (\ref{conforzeta}))). These rather strict conditions on the transformation parameters still allow non-trivial solution if the Grassmann algebra, to which the classical dynamical variables belong, has sufficient generators.

We emphasize that we are maintaining the complex conjugation relation between dotted and undotted spinors. Further, by solving the quadratic equations for $k=1$, we encounter equations for the supergravity transformation parameters, (\ref{confor2}), that generalize those of the supersymmetric background $\mathbb{R}\times S^3$ considered in the literature to arbitrary radius (although we set the auxiliary fields to zero).

A feature of the first FLRW supermultiplet described in this work is the induced fixing of the time gauge $N=a$ from consistency of their supergravity transformation rules. While the conformal time gauge $N=a$ is not an issue, it is desirable to leave the gauge fields arbitrary to use them as Lagrange multipliers in a canonical formulation. Thus, inspired by earlier works, we consider the spatial projection of the quadratic isometry equations. In other words, we are allowing deviations of the FLRW form that are proportional to gauge fields. Doing this allows us to have $\psi_0$ and $\psi_i$ independent of each other, but we can fix them to recover the previous solution. In this sense, the extended multiplet can be considered of the FLRW type. Finally, we comment of the relation with the solution of D'Eath and Hughes, obtained through an ``FRW ansatz", in which, by allowing complex geometry and liberating spinors from the complex conjugation relation, a different constraint can be taken to preserve the vanishing spin-$\frac{3}{2}$ component.

There are several aspects that will be addressed in future work. In particular, the introduction of matter, which will allow us to address the issue of spontaneous supersymmetry breaking and its connection to the scalar potential. We remark that our FLRW reduction of supergravity does not contradict the well-known fact that FLRW backgrounds do not admit Killing spinors. In our case, the background Rarita-Schwinger field does not vanish and neither $\delta_\zeta \psi_m=0$. Instead, the contribution to the spin-$\frac{3}{2}$ component, $\delta_\zeta W_{\be \al \dga}$, is required to vanish. With matter present, we will also be able to identify a mass parameter $m_{1/2}$ for the Weyl spinor $\phi$, the superpartner of the scale factor in the lines of \cite{Tkach1997,Tkach1999}.

Other interesting aspects to be considered include a more elegant gauge-independent definition of FLRW supergravity configurations and an enhanced formulation of the quadratic isometry equations that yields automatically to a reduced supermultiplet, instead of enforcing it by hand.

\appendix

\section{Notation}\label{appendix1}
The index structure of the Lorentz generators are $\sigma^a=\sigma^a_{\al \dal}$, $\bsig^a=\bsig^{a \dal \al}$, where $\sig^{\tilde{0}}=-(\bm 1)_{2\times 2}=\bsig^{\tilde{0}}$, and $\sig^{\tilde{k}}=-\bsig^{\tilde{k}}$ are the standard Pauli matrices. On the other hand,  the bi-spinor form of the tetrad is
\begin{align}\label{sigm}
	\sig^m_{\al \dal}=\sig^a_{\al \dal} e_a^{\ m}, && \bar{\sigma}_m^{\dot{\alpha} \alpha}\equiv e_m^{\ a} \bar{\sigma}_a^{\dot{\alpha} \alpha}.
\end{align}
We use tildes for values of $\sigma^a$ to distinguish them from $\sig^m$: $\sig^a=(\sig^{\tilde{0}},\sig^{\tilde{1}},\sig^{\tilde{2}},\sig^{\tilde{3}})$ $\sig^m=(\sig^0,\sig^1,\sig^2,\sig^3)$.

Spinor indices are raised/lowered with the antisymmetric tensors $(\epsilon^{\al \be}, \ep^{\dal \dbe})\equiv i \sig^{\tilde{2}}$, $(\ep_{\al \be}, \ep_{\dal \dbe})\equiv -i \sig^{\tilde{2}}$,  according to $\chi^\al=\ep^{\al \be} \chi_\be$, $\chi_\al=\ep_{\al \be} \chi^\be$, $\chi^\dal=\ep^{\dal \dbe} \chi_\dbe$, $\chi_\dal=\ep_{\dal \dbe} \chi^\dbe$.

Lorentz generators in the dotted/undotted spinor and 4-vector representations are related by
\begin{align}\label{gens}
	L_\beta^{\ \alpha}=-\frac{1}{2} (\sig^{ba})_{\beta}^{\ \alpha} L_{ba}, &&	L^\dbe_{\ \dal}=-\frac{1}{2} (\bsig^{ba})^\dbe_{\ \dal} L_{ba},
\end{align}
%L_{ba}&=L_\al^{\ \be}(\sig_{ba})_\be^{\ \al}+(\bsig_{ba})^\dal_{\ \dbe} L^\dbe_{\ \dal}
where $\sig^{ba}=\frac{1}{4} (\sig^b \bsig^a-\sig^a \bsig^b)$ and $\bsig^{ba}=\frac{1}{4} (\bsig^b \sig^a-\bsig^a \sig^b)$.

Spinor components are anticommuting elements of a Grassmann algebra \cite{henneaux}: $\chi_\al \eta_\be+\eta_\be \chi_\al=0$, $\bar{\chi}_\dal \bet_\dbe+\bet_\dbe \bar{\chi}_\dal=0$. Thus, contractions satisfy $\eta \chi \equiv \eta^\al \chi_\al=-\chi_\al \eta^\al=\chi^\al \eta_\al=\chi \eta$ and $\bet \bar{\chi} \equiv \bet_\dal \bar{\chi}^\dal=-\bar{\chi}^\dal \bet_\dal=\bar{\chi}_\dal \bet^\dal =\bar{\chi} \bet$.

The torsionless connection and the contorsion are 
\begin{subequations}\label{spinconnection}
	\begin{align}
		2 \omega_{nml}&=e_{na} (\partial_m e_l^{\ a}-\partial_l e_m^{\ a})-e_{la} (\partial_n e_m^{\ a}-\partial_m e_n^{\ a})-e_{ma} (\partial_l e_n^{\ a}-\partial_n e_l^{\ a}), \label{connect} \\
		%%%%%%%%%%%%%%%
		4 \kappa_{nml}&=i (\psi_l \sigma_n \bar{\psi}_m-\psi_m \sigma_n \bar{\psi}_l)-i (\psi_m \sigma_l \bar{\psi}_n-\psi_n \sigma_l \bar{\psi}_m)-i (\psi_n \sigma_m \bar{\psi}_l-\psi_l \sigma_m \bar{\psi}_n), \label{contorsion}
	\end{align}
\end{subequations}
using the curved space $\sig$-matrices (\ref{sigm}).

\subsection{Torsion values}
The torsion components are given by
\begin{align}\label{torsionr}
	T_{NM}^{\ \ \ A}(z)=&\partial_N E_M^{\ A}+(-)^{n(b+m)} E_M^{\ B} \phi_{NB}^{\ \ \ A}-(-)^{nm} \partial_M E_N^{\ A}-(-)^{mb} E_N^{\ B} \phi_{MB}^{\ \ \ A}.
\end{align} 

The torsion components are subject to the Bianchi identities and to additional covariant constraints  \cite{wessbagger}
\begin{subequations}\label{const}
\begin{align}
T_{\underline{\alpha} \underline{\ga}}^{\ \ \ \underline{\gamma}}(z)=T_{ab}^{\ \ c}(z)&=0, \\
T_{\alpha \ga}^{\ \ c}(z)=T_{\dot{\alpha} \dot{\ga}}^{\ \ c}(z)=T_{\underline{\alpha} b}^{\ \ c}(z)=T_{a \underline{\ga} }^{\ \ c}(z)&=0,  \\
T_{\alpha \dot{\ga}}^{\ \ c}(z)=T_{\dot{\ga} \alpha}^{\ \ c}(z)&=2i\, \sigma_{\alpha \dot{\ga}}^{c}.
\end{align}
\end{subequations}
with underline denoting dotted and undotted indices. Some torsion values referred to in this work are
\begin{subequations}\label{torsion}
	\begin{align}
		T_{bm}^{\ \ \ a}|&=0, \\
		T_{nm}^{\ \ \ a}|&=\p_n e_m^{\ a}-\p_n e_m^{\ a}+\phi_{nm}^{\ \ \ a}-\phi_{mn}^{\ \ \ a}=\frac{i}{2} (\psi_n \sig^a \bar{\psi}_m-\psi_m \sig^a \bar{\psi}_n), \label{torsion1}\\ 
		T_{b m}^{\ \ \ \al}|&=e_b^{\ n} T_{nm}^{\ \ \ \al}|-\tf \psi_b^{\ \la} T_{\la m}^{\ \ \ \al}|-\tf \bar{\psi}_{b\dot{\la}} T_{\ \ m}^{\dot{\la} \ \al}|, \\
		2 T_{nm}^{\ \ \ \al}|&=\partial_n \psi_m^{\ \al}+\psi_m^{\ \be} \phi_{n \be}^{\ \ \ \al}-\psi_n^{\ \be} \phi_{m \be}^{\ \ \ \al}-\partial_m \psi_n^{\ \al}, \label{ftor}
	\end{align}
\end{subequations}
where $T_{\la m}^{\ \ \ \al}|$ and $T_{\ \ m}^{\dot{\la} \ \al}|$ depend on auxiliary fields \cite{wessbagger}.

\section{Spatially flat conformal supergravity in spherical coordinates}\label{sphere}
Using the diagonal tetrad $\bm E^{\tilde{0}}=-N dt$, $\bm E^{\tilde{1}}=a dr$, $\bm E^{\tilde{2}}=a r d\theta$, $\bm E^{\tilde{3}}=a r \sin \theta d \varphi$, we get the connection components (\ref{conflatsphere})
\begin{subequations}\label{conflatsphere}
	\begin{align}
		\omega_{r\beta}^{\ \ \alpha}&=\frac{\dot{a}}{2 N}  (\sigma^{\tilde{0}} \bar{\sigma}^{\tilde{1}})_\beta^{\ \alpha}, \\
		%%%%%%%%%%%%%%%%%%%%%%%%%
		\omega_{\theta \beta}^{\ \ \ \alpha}&=\frac{\dot{a}}{2N} r (\sigma^{\tilde{0}} \bar{\sigma}^{\tilde{2}})_\beta^{\ \alpha}+\frac{i}{2} (\sigma^{\tilde{0}} \bar{\sigma}^{\tilde{3}})_\beta^{\ \alpha}, \\
		%%%%%%%%%%%%%%%%%%%%%%
		\omega_{\varphi \beta}^{\ \ \ \alpha}&=\frac{\dot{a}}{2 N} r \sin \theta (\sigma^{\tilde{0}} \bar{\sigma}^{\tilde{3}})_\beta^{\ \alpha}-\frac{i}{2} \sin \theta (\sigma^{\tilde{0}} \bar{\sigma}^{\tilde{2}})_\beta^{\ \alpha}+\frac{i}{2} \cos \theta (\sigma^{\tilde{0}} \bar{\sigma}^{\tilde{1}})_\beta^{\ \alpha}.
	\end{align}
\end{subequations}

The vanishing time-space components $V_{0i}$  yield the following equations
\begin{subequations}\label{constric}
	\begin{align}
0=&\frac{1}{N}\Big(\frac{\dot{a}}{2 a} \zeta-\partial_t \zeta\Big) \sigma^{\tilde{1}}\bar{\phi}-\frac{1}{a} (\partial_r \zeta) \sigma^{\tilde{0}} \bar{\phi}, \\
		%%%%%%%%%%%%%%%%%%%%%%%%%%
0=&\frac{1}{N} \Big(\frac{\dot{a}}{2 a} \zeta-\partial_t \zeta\Big) \sigma^{\tilde{2}} \bar{\phi}+\frac{1}{r a} \Big(\frac{1}{2} \zeta \sigma^{\tilde{1}} \bar{\sigma}^{\tilde{2}}-\partial_\theta \zeta^\alpha\Big) \sigma^{\tilde{0}} \bar{\phi}, \\
		%%%%%%%%%%%%%%%%%%%%%%%%%%
0=&\frac{1}{N} \Big(\frac{\dot{a}}{2 a} \zeta-\partial_t \zeta\Big) \sigma^{\tilde{3}} \bar{\phi}+\frac{1}{r a \sin \theta} \Big(\frac{\sin \theta}{2} \zeta \sigma^{\tilde{1}} \bar{\sigma}^{\tilde{3}}+\frac{\cos \theta}{2} \zeta \sigma^{\tilde{2}} \bar{\sigma}^{\tilde{3}}-\partial_\phi \zeta\Big) \sigma^{\tilde{0}} \bar{\phi}.
	\end{align}
\end{subequations}
Substituting the ansatz $\zeta^\alpha(t,\theta,\varphi)=\zeta'^\beta(t) \Lambda_\beta^{\ \alpha}(\theta,\varphi)$ into (\ref{constric}) yields
\begin{subequations}\label{conditions}
	\begin{align}
		0=&\dot{\zeta}'-\frac{\dot{a}}{2a} \zeta', \label{timezeta} \\ 
		0=&\partial_\theta \Lambda_\beta^{\ \alpha}-\frac{1}{2} \Lambda_\beta^{\ \gamma} (\sigma^{\tilde{1}} \bar{\sigma}^{\tilde{2}})_\gamma^{\ \alpha}, \\
		0=&\partial_\phi \Lambda-\frac{\sin \theta}{2} \Lambda \sigma^{\tilde{1}} \bar{\sigma}^{\tilde{3}}-\frac{\cos \theta}{2} \Lambda \sigma^{\tilde{2}} \bar{\sigma}^{\tilde{3}}.
	\end{align}
\end{subequations}
A particular solution for $\Lambda$ is the spatial rotation
\begin{align}\label{lor}
	\Lambda(\theta,\varphi)=\frac{1}{\sqrt{2}} \begin{bmatrix}
		-\sqrt{i} e^{-i (\theta+\varphi)/2} & -\sqrt{i} e^{i (\theta-\varphi)/2} \\
		-i \sqrt{i} e^{-i (\theta-\varphi)/2} & i \sqrt{i} e^{i (\theta+\varphi)/2}
	\end{bmatrix}.
\end{align}

\centerline{\bf Acknowledgements}
N.E. Martínez-Pérez thanks SECIHTI for financial support. We thank the anonymous referee for useful comments.

% BibTeX users please use
% \bibliographystyle{plain}
 \bibliographystyle{unsrt}
 \bibliography{isometries}
%
% Non-BibTeX users please use
%\begin{thebibliography}{}
%
% and use \bibitem to create references.
%
%\bibitem{RefJ}
% Format for Journal Reference
%Author, Journal \textbf{Volume}, (year) page numbers.
% Format for books
%\bibitem{RefB}
%Author, \textit{Book title} (Publisher, place year) page numbers
% etc
%\end{thebibliography}

\end{document}